\documentclass[final,5p,times,twocolumn]{elsarticle}
\usepackage{graphicx}
\usepackage{amssymb}
\usepackage{lineno}
\usepackage{amssymb}
\usepackage[colorlinks,linkcolor=SB,citecolor=red]{hyperref}
\usepackage{geometry}
\usepackage{graphicx}
\usepackage{natbib}
\usepackage{etoolbox}
\usepackage{float}

\usepackage{hyperref}
\usepackage{float}
\usepackage[caption = false]{subfig}
\usepackage{amsmath,amssymb,amsfonts}

\apptocmd{\sloppy}{\hbadness 10000\relax}{}{}

\biboptions{comma,sort&compress}
\journal{}

\begin{document}

\begin{frontmatter}


\title{Mapping Coupled Time-series Onto Complex Network}



\author[ardalankiaaddress1,ccnsdaddress]{Jamshid Ardalankia}
\author[grjafariaddress1]{Jafar Askari}
\author[Sheykhaliaddress1,Sheykhaliaddress2]{Somaye Sheykhali}
\author[Havenaddress]{Emmanuel Haven}
\ead{ehaven@mun.ca}
\author[ccnsdaddress,grjafariaddress1,grjafariaddress3]{G.Reza Jafari}
\ead{gjafari@gmail.com}

\address[ardalankiaaddress1]{Department of Financial Management, Shahid Beheshti University, G.C., Evin, Tehran, 19839, Iran}
\address[ccnsdaddress]{Center for Complex Networks and Social Data Science, Department of Physics, Shahid Beheshti University, G.C., Evin, Tehran, 19839, Iran}
\address[grjafariaddress1]{Department of Physics, Shahid Beheshti University, G.C., Evin, Tehran, 19839, Iran}
\address[Sheykhaliaddress1]{Department of Physics, University of Zanjan (ZNU), Zanjan, 45371-38791, Iran}
\address[Sheykhaliaddress2]{Instituto de Física Interdisciplinary Sistemas Complejos IFISC (CSIC-UIB), Palma de Mallorca, E07122, Spain}
\address[Havenaddress]{Faculty of Business Administration, Memorial University, St. John's, Canada and IQSCS, UK}
\address[grjafariaddress3]{Department of Network and Data Science, Central European University, 1051 Budapest, Hungary}

\begin{abstract}
In order to extract hidden joint information from two possibly uncorrelated time-series, we explored the measures of network science. Alongside common methods in time-series analysis of the economic markets, mapping the joint structure of two time-series onto a network provides insight into hidden aspects embedded in the couplings. We discretize the amplitude of two time-series and investigate relative simultaneous locations of those amplitudes. Each segment of a discretized amplitude is considered as a node. The simultaneity of the amplitudes of the two time-series is considered as the edges in the network. The frequency of occurrences forms the weighted edges. In order to extract information, we need to measure that to what extent the coupling deviates from the coupling of two uncoupled series. Also, we need to measure that to what extent the couplings inherit their characteristics from a Gaussian distribution or a non-Gaussian distribution. We mapped the network from two surrogate time-series. The results show that the couplings of markets possess some features which diverge from the same features of the network mapped from white noise, and from the network mapped from two surrogate time-series. These deviations prove that there exist joint information and cross-correlation therein. By applying the network's topological and statistical measures and the deformation ratio in the joint probability distribution, we distinguished basic structures of cross-correlation and coupling of cross-markets. It was discovered that even two possibly known uncorrelated markets may possess some joint patterns with each other. Thereby, those markets should be examined as coupled and \textit{weakly} coupled markets.

\end{abstract}


\end{frontmatter}


\section{Introduction}
It is intriguing to study the joint information of two time-series by
mapping their coupling onto a network. Several added advantages appear if
one investigates the two time-series after applying the surrogate method,
and to then find the joint structures of those two time-series by mapping
them onto a network. By this approach, the sources of coupled structures are
revealed (correlation and fat-tailed distribution). The reasoning behind
applying this procedure provides from the numerous measures in network
science~\cite{Barabsi1999,Newman2003,Newman2006,NewmanGirvan2004}. In order
to extract more hidden information from time series, network science has
been successfully utilized for analyzing the extracted information from time
series coming from a wide variety of fields, all through the analysis of the
derived network~\cite%
{Zhang2017,Zou2019,Sun2014,lacasa2008time,Shirazi2009,Campanharo2011,Zhang2019,Jacob2019}%
.\newline
We investigate the coupling and cross-correlation of three financial
time-series, DJIA, S\&P500 and SSEC. As mentioned in~\cite{Zhang2017}, stock
markets can be characterized as systems with joint structures and
simultaneous behaviors and they can be analyzed by mapping their joint
structures onto a network. Zhang~\textit{etal}~\cite{Zhang2017} introduced ordered
patterns for some chaotic time-series. They mapped the evolution of patterns
onto a network. This transition network formed a chain with forbidden
patterns. The weighted edges correspond to the frequency of patterns. It was
shown that by changing the parameters of the system dynamics in ordinal
transition networks, the temporal evolution and forbidden phases in the
network may change. Also, by mapping the trajectories and lagged effects of
a dynamical system onto a network, one can form a Markov chain~\cite%
{McCullough2015}. Some other methods have been applied to map time-series
analysis to network science. For instance, visibility graphs~\cite%
{lacasa2008time,Xiong2019,Luque2009HorizontalVG,Stephen2015,Bianchi2017}
reveal the structural topology of networks of coupled time-series in whether
the time-series are random, periodic (ordered) or fractal. This will cause
the mapping algorithm to result in respectively, a random
network, regular network and scale-free network~\cite{lacasa2008time}.
Mapping time-series onto a network is also employed by a visibility graph
algorithm to assess network topological and statistical measures of
financial markets with different Hurst exponents~\cite{Stephen2015}. Mapping
multi-variate time-series also result in multi-layer networks~\cite%
{Lacasa2015,boccaletti2014structure}. The visibility graph method for
mapping time-series onto a multi-layer network, has a wide applicability in
machine learning~\cite{Bianchi2017}. A multi-scale mapping of time-series
onto a network and the transmission of ordinal regression patterns between
two time-series in the local trends of non-stationary time-series, provide
for useful results too~\cite{Gao2014}. With respect to the cross-correlation
networks~\cite{Mehraban2013}, the network properties such as the clustering
coefficient, the efficiency, the cross-correlation degree of
cross-correlation interval and also the modularity of dynamic states, have
all been investigated~\cite{Feng2017}. An example of another real-world
application provides from the mapping of time-series onto a network in
tourism management~\cite{Baggio2016}. \\
One of the main strengths of network science in dynamic systems can be found
in higher-order analysis~\cite{Scholtes2017controllability,Lambiotte2019}.
Network analysis is of tremendous use in finance and economics. It helps
scholars to explore deeper systemic risk evaluations~\cite%
{Battiston2016,Perillo2018,Amini2010,Shirazi2017,Kaushik2013,Habibnia2017}.
We will compare
the patterns in mapped networks of two market based time-series with the
patterns in mapped networks with fractional Gaussian noises (fGns). FGns are
known as specific random series with the range of anti-persistent, white
noise, and persistent behavior where the so called Hurst exponent has
relevance. The Hurst exponent is a criterion which informs to what extent
two time-series are coupled in various time-scales~\cite%
{Ardalankia2020,Caraiani2015,Hedayatifar2011}. The pattern of some
measurements in network science shows that there is coupled information
embedded in the joint systems. Some measurements are significantly close to
the networks mapped from fGns. However, there exist measurements where none
of the networks converge to a definite value. Based on the segregation of
those networks, the information transitions and measurements with closer
values are revealed. The coupling and cross-correlation in financial
time-series intrinsically contain scaling behaviors~\cite%
{Ardalankia2020,Caraiani2015,Hedayatifar2011}. Those scaling behaviours not
only emerge in temporal aspects, but they also appear in higher statistical
moments of price return distributions. In this context, the present study
casts light into the behavior of the couplings between financial time-series
by applying novel measures of network science.\newline
We are supposed to capture temporal/dynamic behaviors of the financial
time-series by mapping onto a network as follows, by:\newline
\textbf{I.} introducing the mapping algorithm from coupled time-series onto
a network;\newline
\textbf{II.} constructing a network obtained by mapping the coupling of two
financial time-series, and;\newline
\textbf{III.} constructing the networks mapped from fractional Gaussian
noises (fGns) which are coupled by their 1-step lag with a range of Hurst
exponents;\newline
\textbf{IV.} comparing the obtained networks and extracting the hidden
features of couplings.

\section{The Mapping Algorithm}

\begin{figure}[h]
	\includegraphics[width=0.4\textwidth]{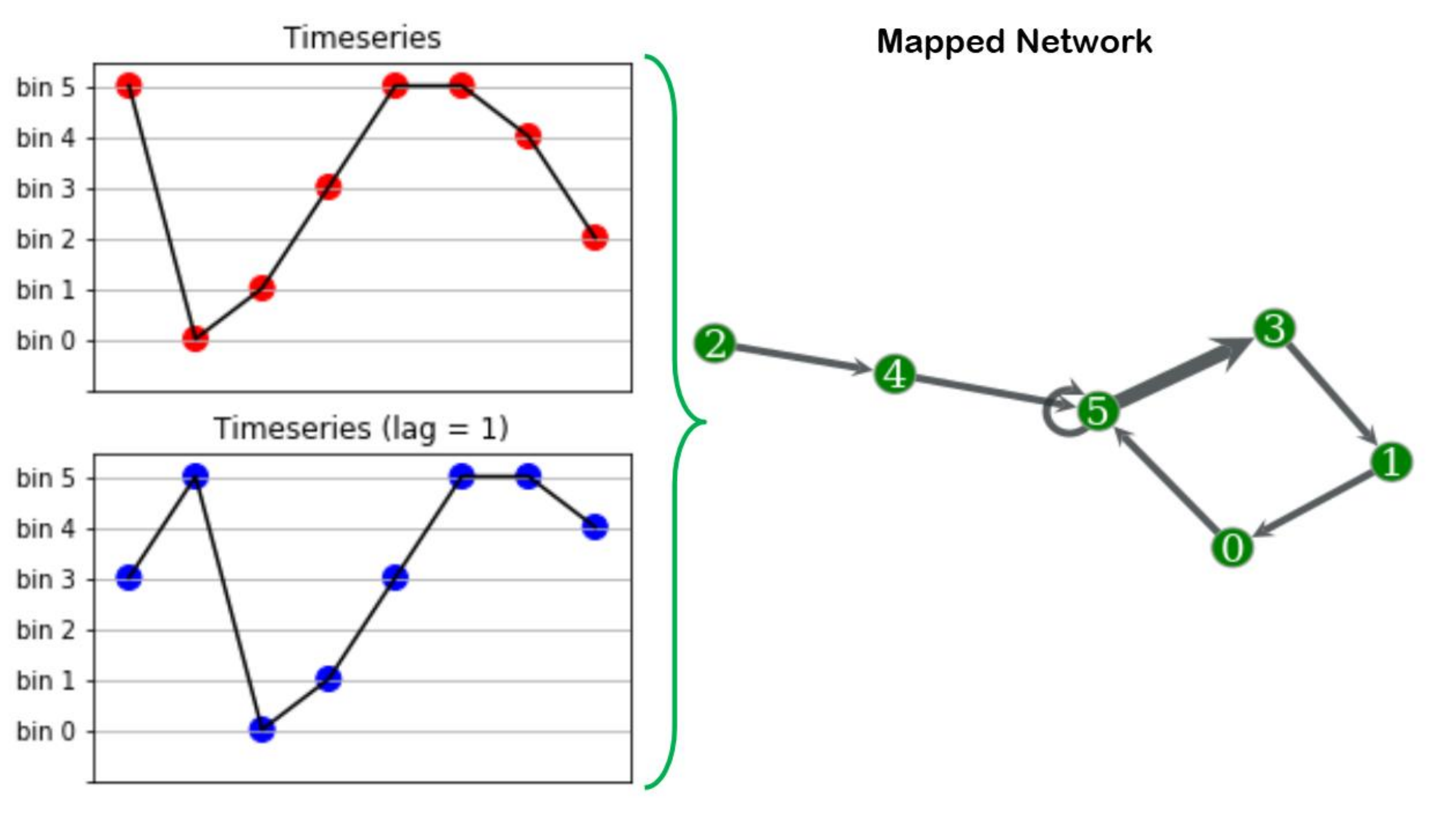}  %
	\includegraphics[width=0.4\textwidth]{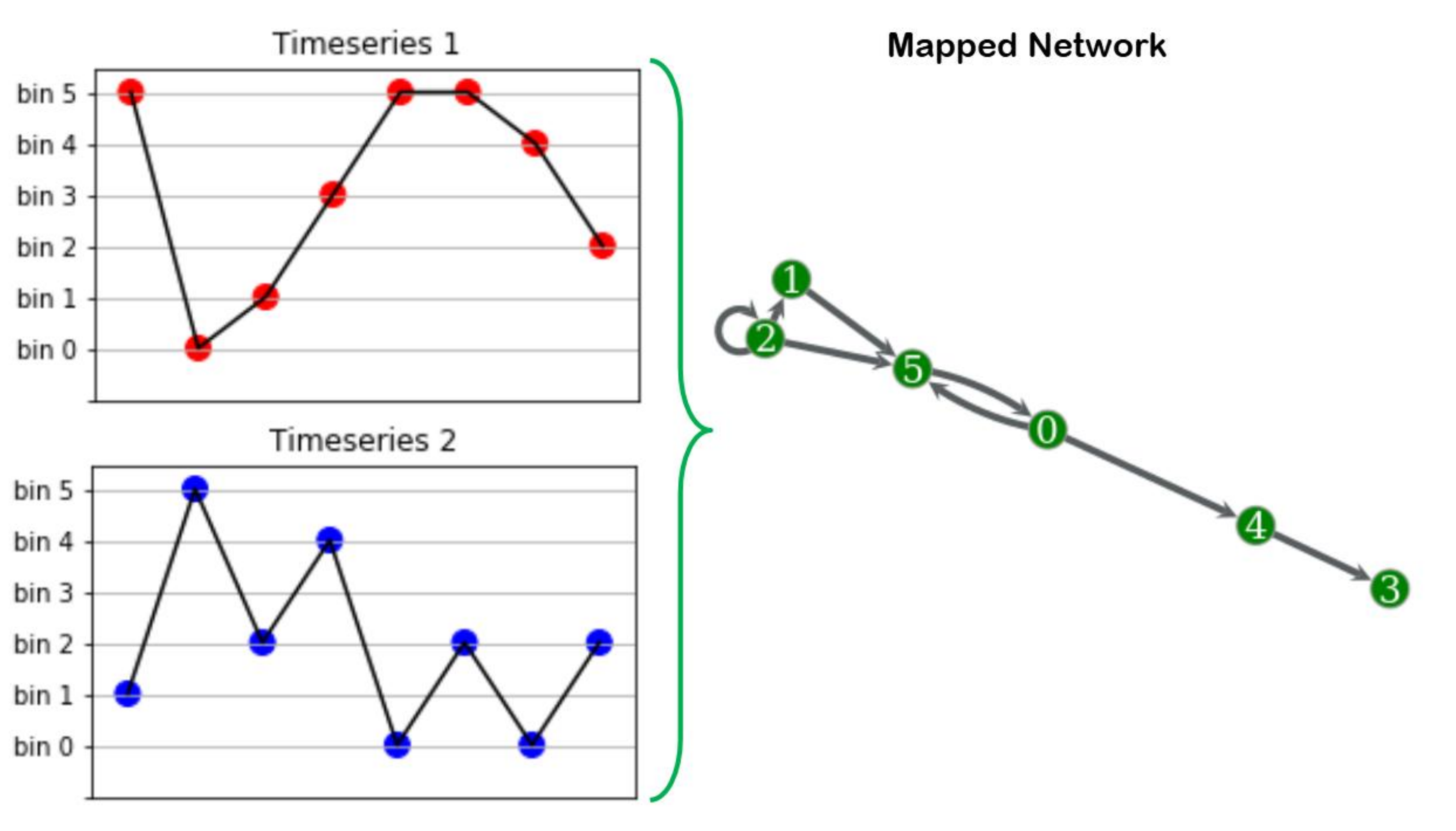}  
	\caption{We demonstrate the mapping algorithm. This figure depicts the way
		links in the network are generated. As shown, when the amplitudes
		corresponding to two time-series are located in the same amplitude-bins
		(nodes), a weighted \textit{self-loop} is considered. However, when the
		amplitudes are not located in the same bin, two nodes connecting with a
		weighted \textit{edge} are generated. The term \textit{weight} implies the
		frequency of this directed situation-- and can be applied with the
		persistence of edges~\protect\cite{Rocha2017}. The outcome will be a
		temporal network.}
	\label{fig_algo}
\end{figure}

As shown in Fig.~\ref{fig_algo}, for two time-series: $X(t)=%
\{x(t_{1}),x(t_{2}),...,x(t_{N})\}$ and, $Y(t)=%
\{y(t_{1}),y(t_{2}),...,y(t_{N})\}$, we have:\newline
I- The joint probability matrix is constructed from these time-series.
	It will be the adjacency matrix of the network. The frequency of the above-mentioned conditions represents 
	the weighted adjacency matrix among the two time-series.\\
	II- The iteration occurs on the data-points $t= 1, 2, ..., N$. The amplitudes of $x(t)$ and $y(t)$ are discretized 
	to an equal number of bins, and each bin is considered as a node in the network.\\
	III- For any $t = t_n$:\\
	\indent If $x(t_n) = y(t_n)=i$: a self-loop for the node (bin) $i$ is constructed.\\
	\indent If $x(t_n) = i$ and, $y(t_n)=j$, $i\ne j$: an edge between nodes $i$ and $j$ is constructed.%
\newline
The number of bins is a matter of trade-off. 
	A high number of bins shortens the width of bins. 
	Hence, extremely narrow bins contribute to noise detection. 
	Conversely, extremely wide bins contribute to extremely low information extraction.
Considering the amplitude-wise scaling features of financial correlations~%
\cite{Ardalankia2020,Caraiani2015,Hedayatifar2011}, alongside the fact that
the correlation coefficients just reveal linear co-behaviors, there exists a
vital need to consider the effects of direction and size of the
fluctuations. Those amplitudes may contain nonlinear behaviors. Hence, in
our work without the need for necessarily linear relations, the couplings
are defined. This procedure can be explored by both temporal-intervals~\cite%
{Feng2017} and amplitude-intervals~\cite{Shirazi2009}. We generate discrete
intervals to evaluate the amplitude of the markets and we then map those
amplitudes (nodes) and their relations (edges) onto a network.

\section{Mapping Single Time-series Onto a Network}

\begin{figure}[t]
	\centering
	\subfloat{{\includegraphics[scale=0.20]{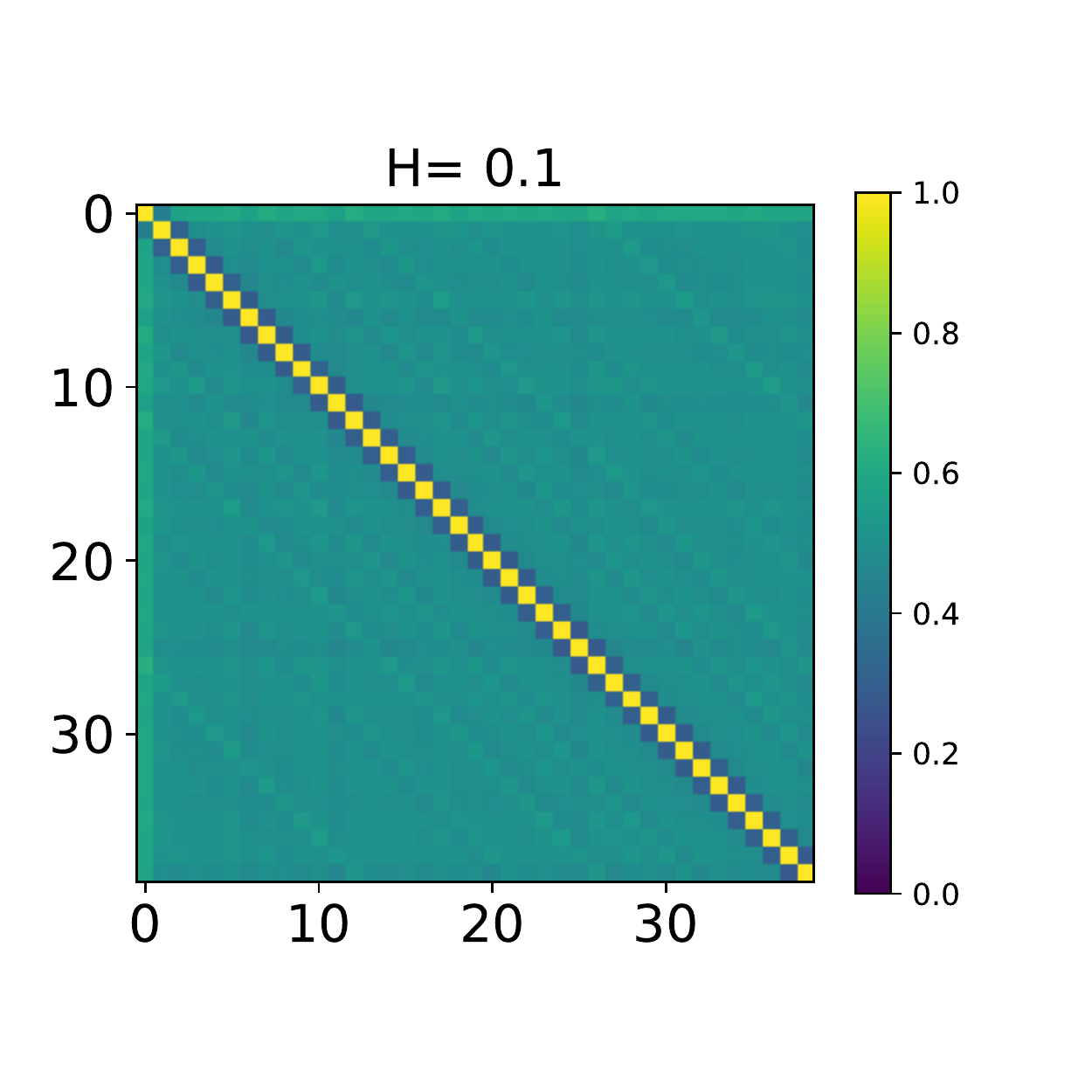}}}  %
	\subfloat{{\includegraphics[scale=0.20]{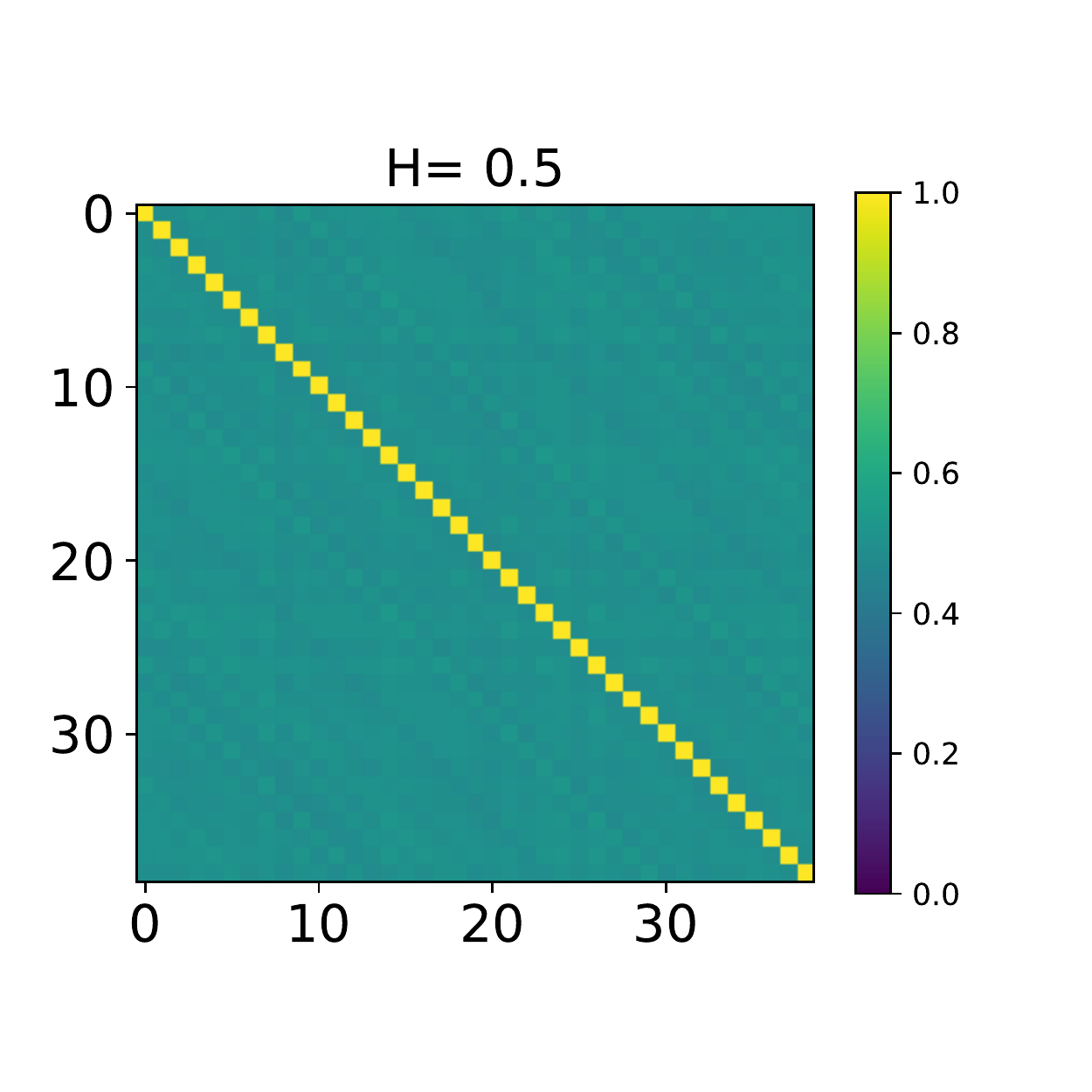}}}  %
	\subfloat{{\includegraphics[scale=0.20]{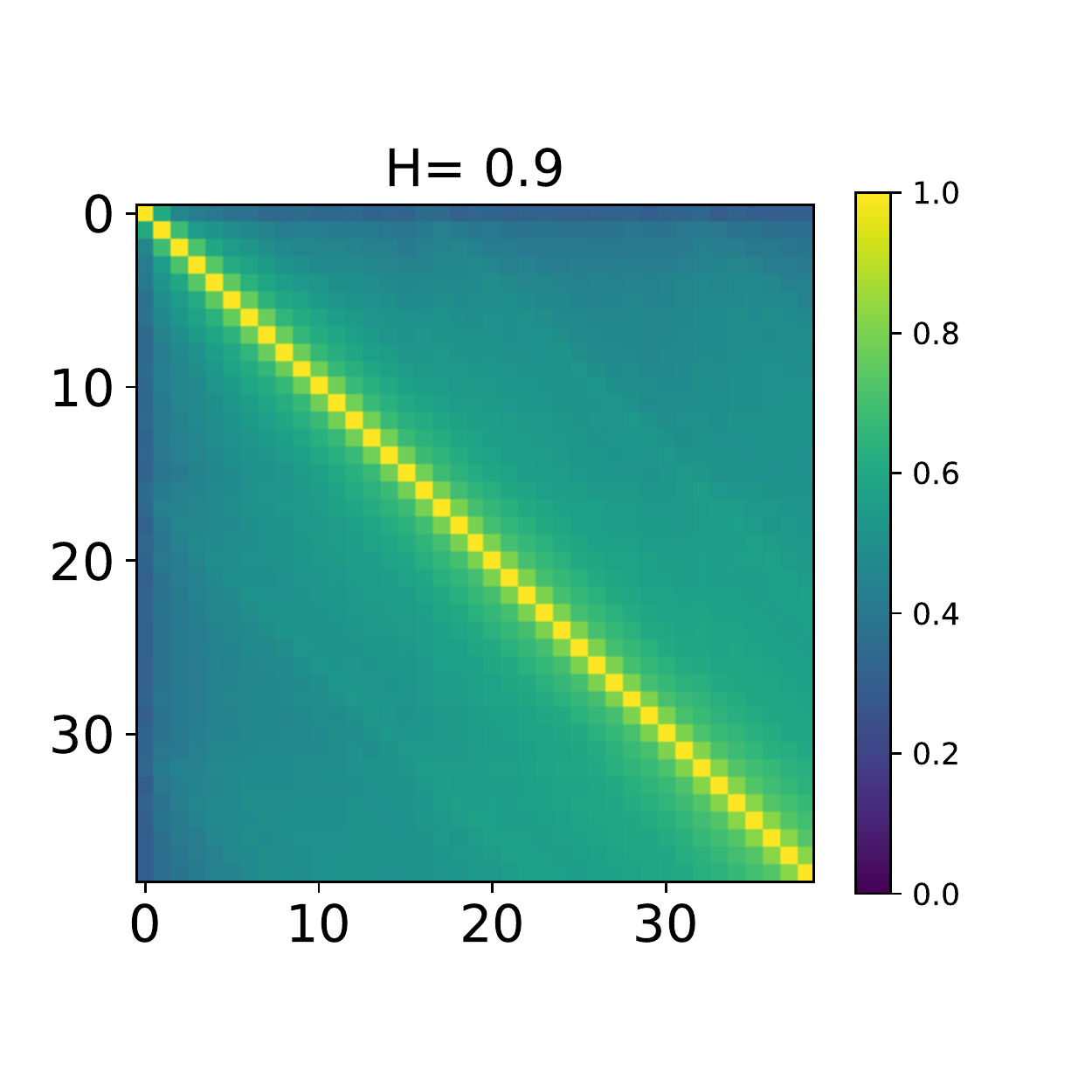}}} 
	\par
	\subfloat{{\includegraphics[scale=0.25]{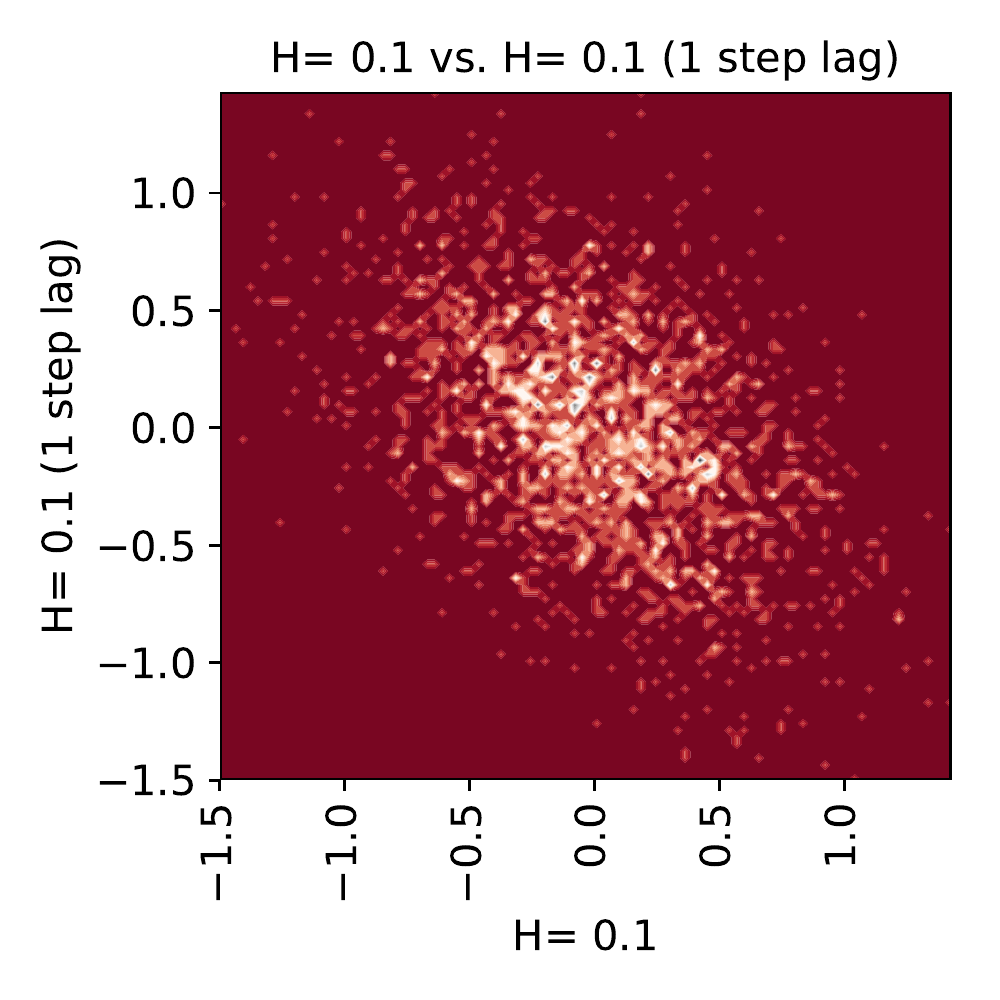}}}  \subfloat{{%
			\includegraphics[scale=0.25]{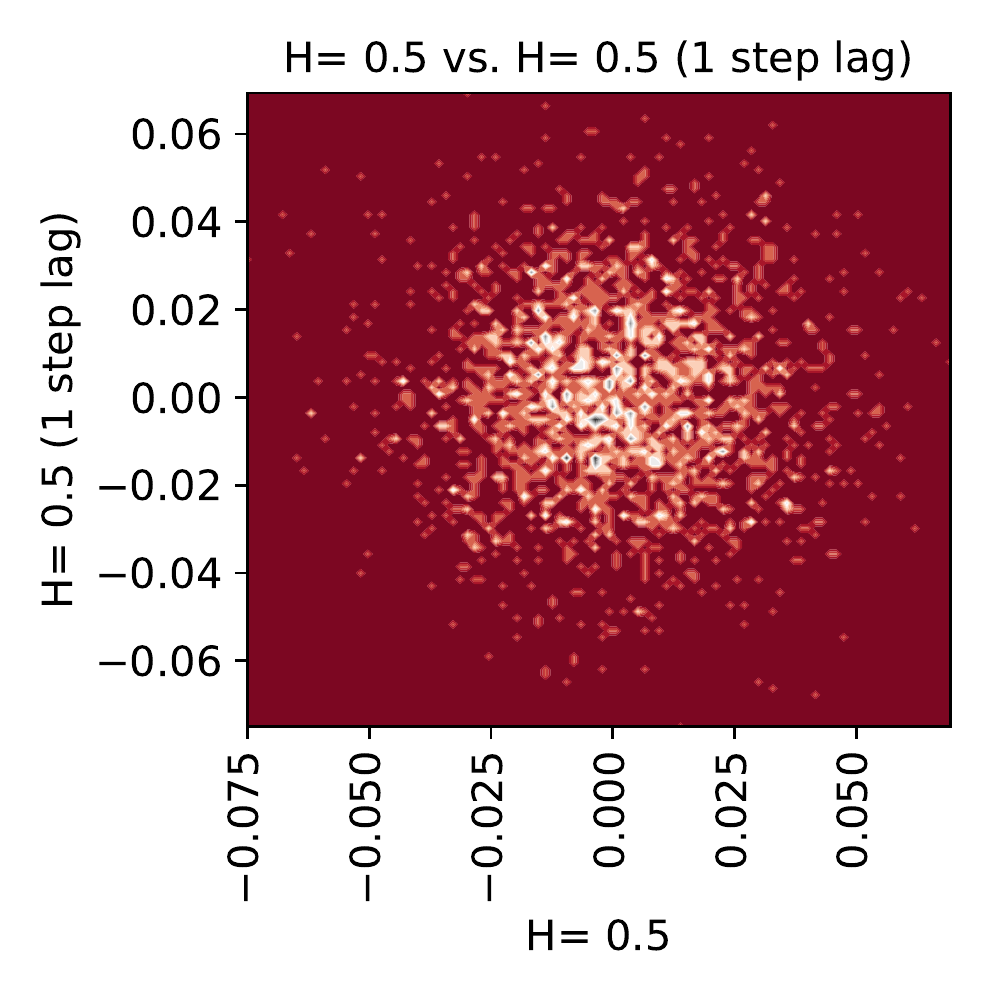}}}  \subfloat{{%
			\includegraphics[scale=0.25]{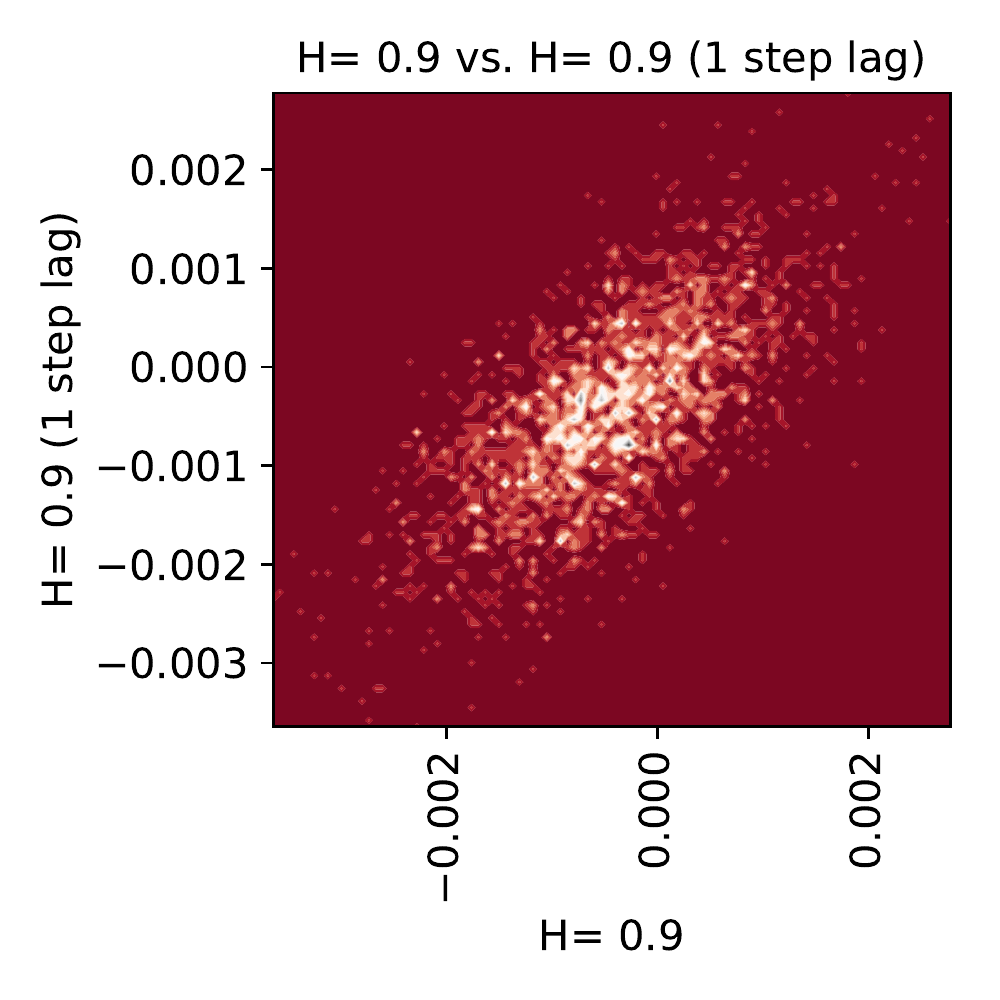}}} 
	\par
	\subfloat{{\includegraphics[scale=0.10]{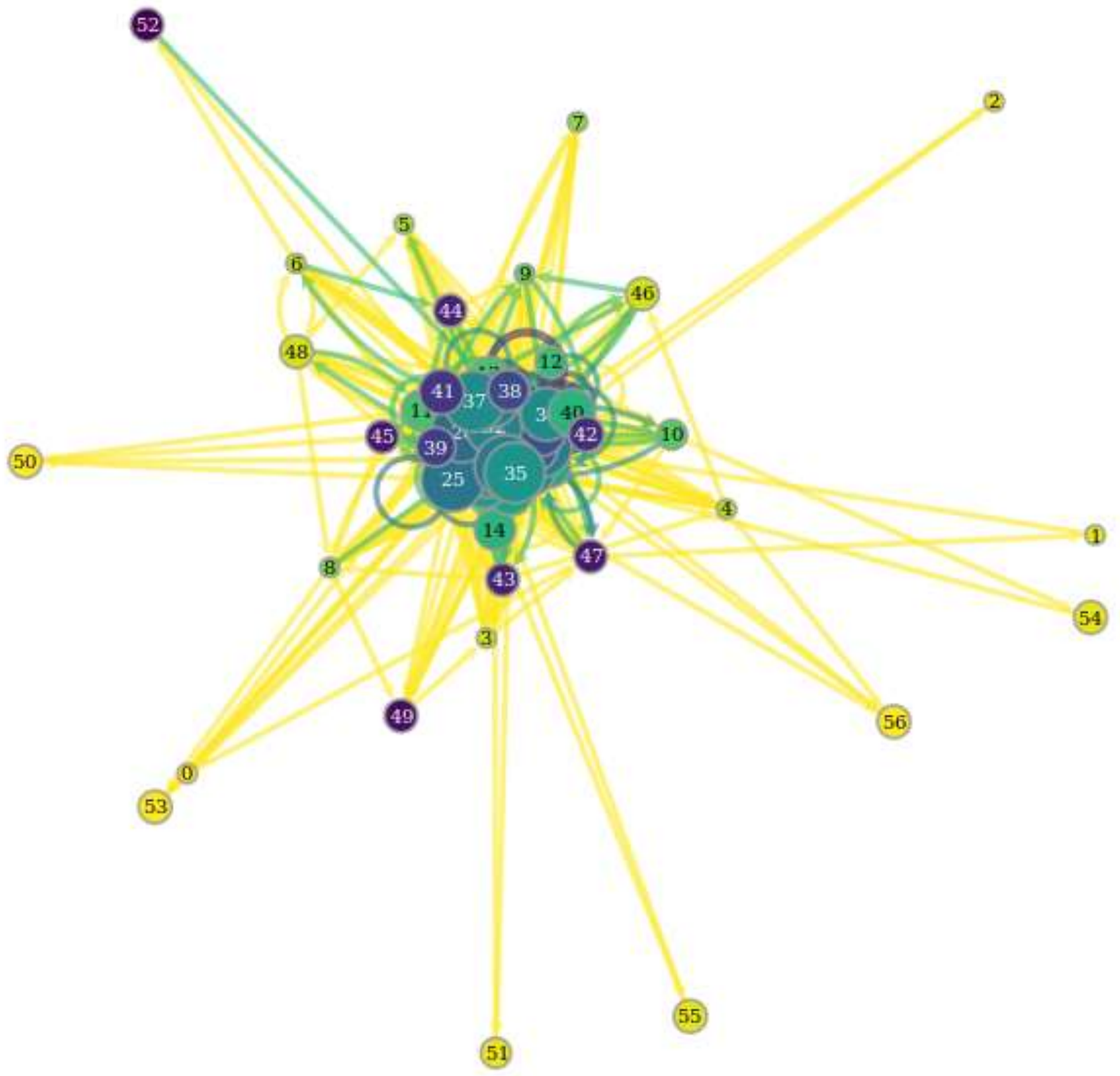}}} \subfloat{{			%
			\includegraphics[scale=0.10]{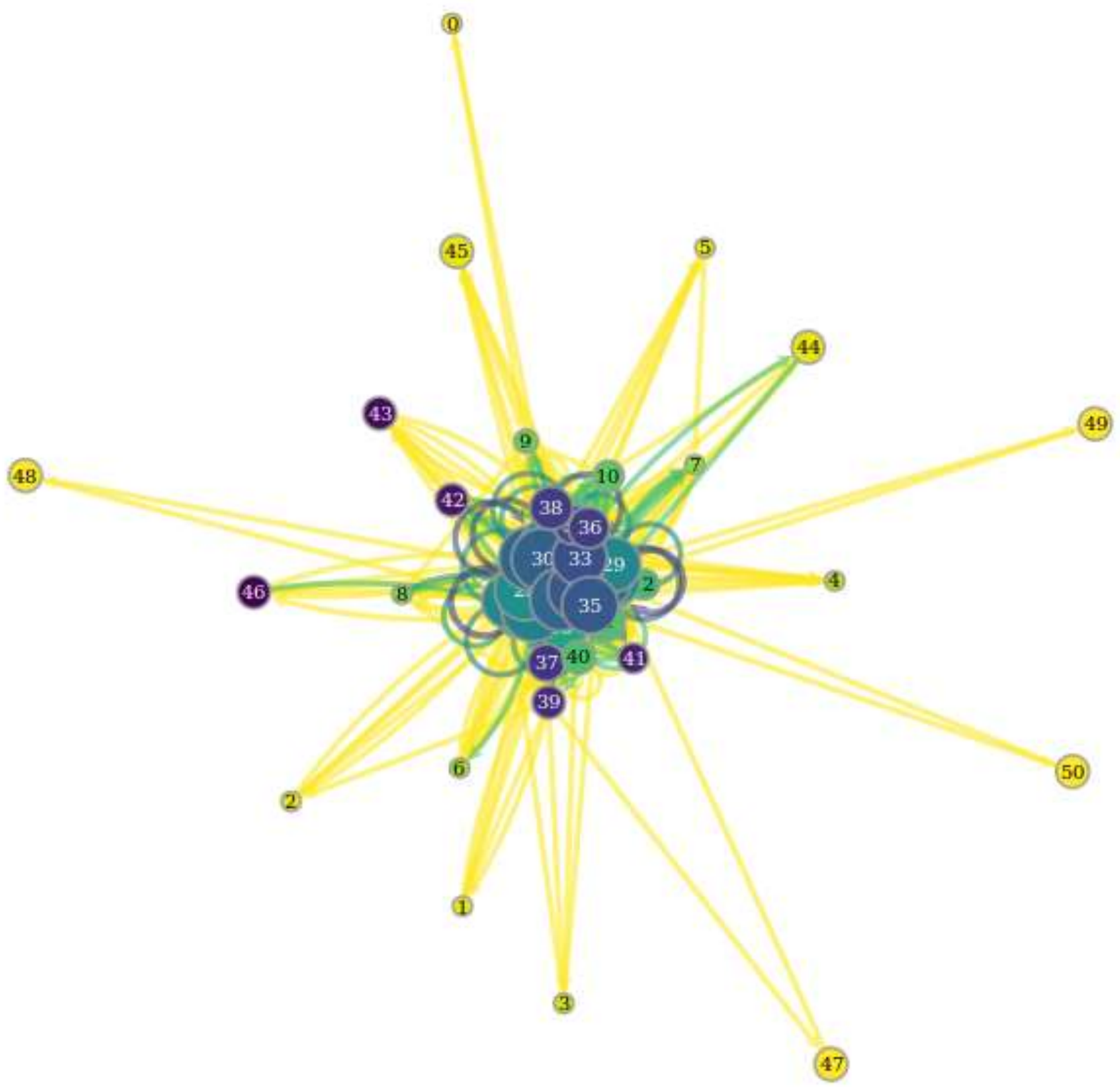}}} \subfloat{{			%
			\includegraphics[scale=0.10]{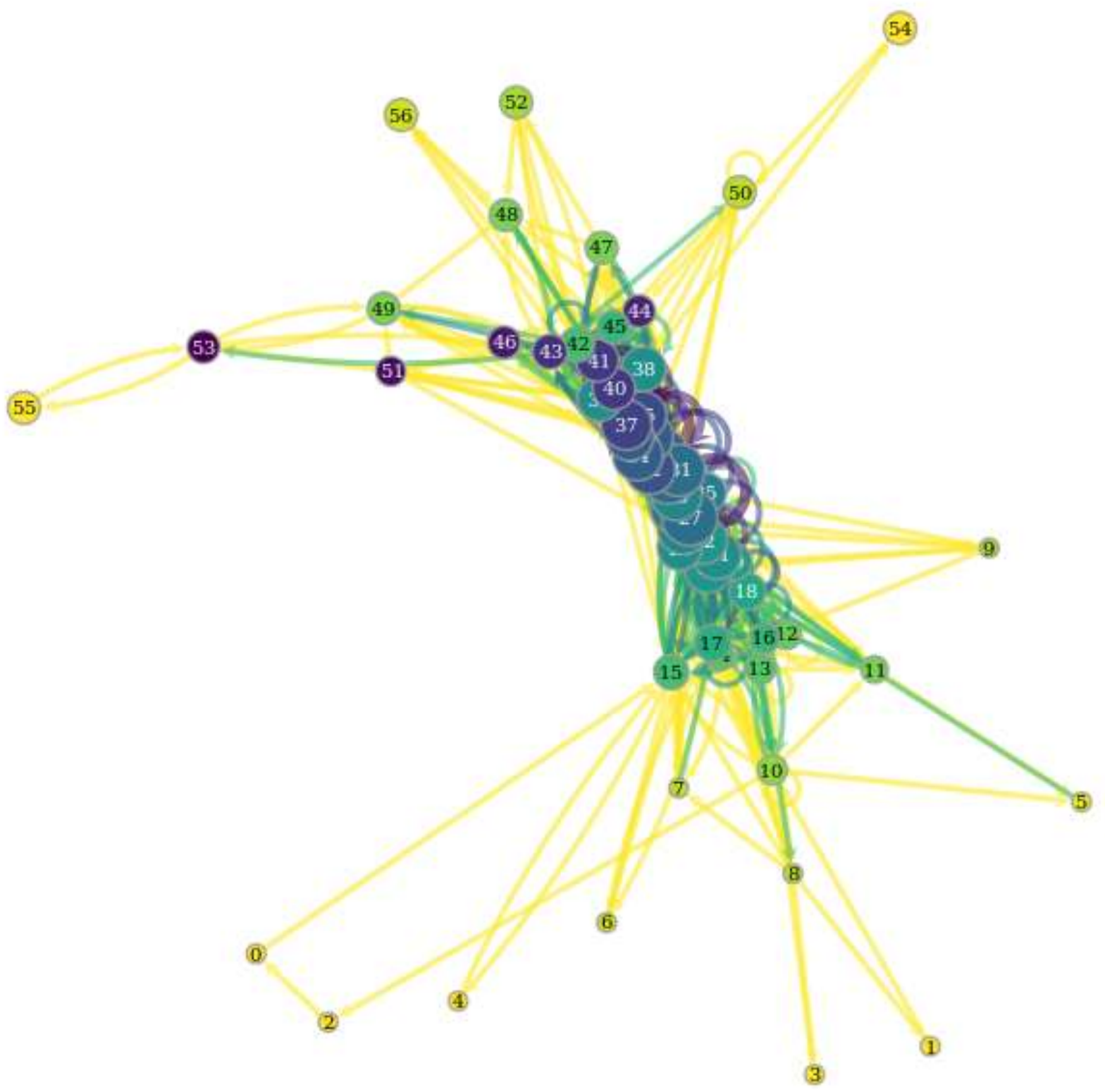}}} 
	\par
	\subfloat{{\includegraphics[scale=0.5]{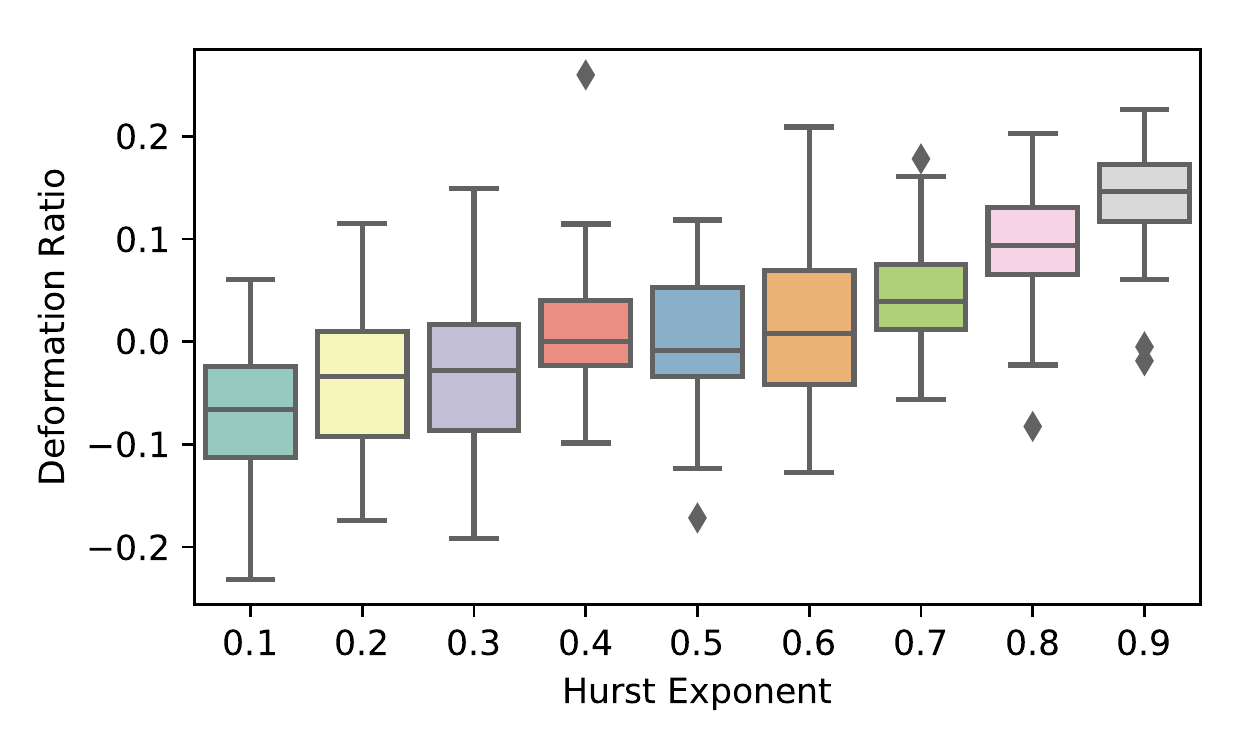}}} 
	\caption{The general temporal pattern of an anti-persistent (Hurst$<$0.5),
		white noise (Hurst=0.5) and persistent (Hurst$>$0.5) systems are demonstrated;
		1st row) The auto-correlation matrices relating to Hurst= 0.1, 0.5, 0.9; 2nd
		row) Adjacency Matrices relating to Hurst= 0.1, 0.5, 0.9; 3rd row). The
		Network's Topology of time-series with Hurst= 0.1, 0.5, 0.9; 4th row)
		Deformation ratio $R$ versus their corresponding Hurst exponents are shown.
		As shown, the general trend of the deformation ratio (in Eq.~\protect\ref%
		{eq1}) versus the Hurst exponent is ascending.}
	\label{network_topology}
\end{figure}

We discretize the amplitudes of a series and also its 1-step-lag series. By
this segmentation and by converting them onto several bins~\cite{Shirazi2009}%
, we couple those amplitudes. We consider these amplitude-bins as nodes in a
network. The top subfigure in Fig.~\ref{fig_algo}, shows the way we design
the algorithm. The Hurst exponent of a system implies how two
time-series --also one single time-series and its lags-- have a coupling in
a persistent (Hurst$>$0.5), white noise (Hurst=0.5), and
anti-persistent (Hurst$<$0.5) manner. This is an intrinsic and a
structural characteristic of developed and emerging financial markets~\cite%
{Ardalankia2020}. Initially, we generate a total of 288 fractional Gaussian
noises (fGns) with Hurst exponents ranging from anti-persistence to
persistence (0.1 to 0.9) and 2000 data-points for each series. A high Hurst
exponent is an identification of stronger coupling. Respecting the shape of
joints and their resulting networks, in Fig.~\ref{network_topology}, it is
depicted that a high Hurst exponent leads to a high elongation around the
main diameter of joint probabilities. To quantify the elongation of
couplings, we introduce a deformation parameter, $R$, based on the standard
deviations along diameters of joint probabilities. This parameter widely
clarifies the couplings behavior, and it is quantified by Eq.~\ref{eq1}; 
\begin{equation}
R=\frac{\sigma _{i}-\sigma _{j}}{max\{\sigma _{i},\sigma _{j}\}};
\label{eq1}
\end{equation}%
%

\begin{figure*}[h]
	\centering
	\includegraphics[width=1\textwidth]{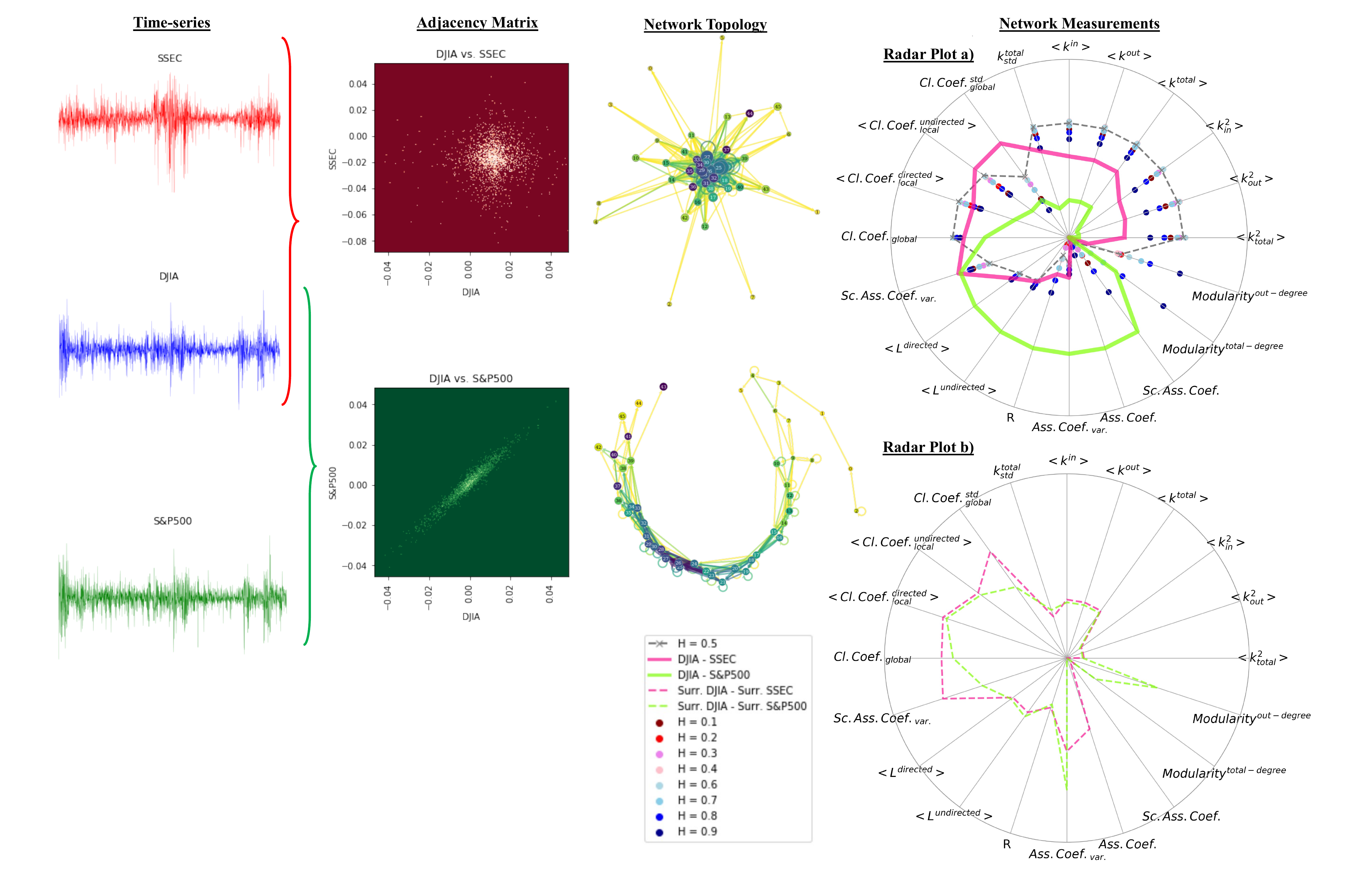}  
	\caption{The mapping of the coupling of time-series onto a network is shown:
		from left, the 1st column shows the time-series of the markets such as SSEC,
		DJIA and S\&P500 in daily resolution during 2000 days until Jul. $31^{st}$
		2019; the 2nd column depicts DJIA-SSEC adjacency matrix and DJIA-S\&P500
		adjacency matrix; the 3rd column illustrates the topology of networks
		corresponding to DJIA-SSEC coupling and DJIA-S\&P500 coupling--simulated by
		Graph-tool~\protect\cite{peixoto_graph-tool_2014}. The radar plots on the
		right side provide fully comparative topological and statistical features of
		the networks. Radar plot (a) shows significant deviating patterns between
		networks mapped from joint series. Since the obtained networks are
		significantly different from each other, it proves that the type of
		couplings are typically different. To consider more extensive test systems,
		in radar plot (b), the market series are surrogated (see Appendix A) and the
		coupling of surrogated series are mapped to the network. The outcome is
		interesting where the patterns have higher conformity rather than the
		original time-series in radar plot (a). Hence, radar plot (b) shows that,
		after surrogate method, the deviations which their sources relate to the
		fat-tailed PDFs, come closer to each other. This means that the deviations,
		with the source of non-Gaussianity, are eliminated and the deviations with
		merely the source of correlation remain.}
	\label{fig_result}
\end{figure*}
\begin{table*}[]
	\centering
	\resizebox{\textwidth}{!}{		\begin{tabular}{cllllllllll}
			\hline
			\multicolumn{1}{|c|}{\textbf{Couplings}}    & \multicolumn{1}{c|}{$<k^2_{total}>$}     & \multicolumn{1}{c|}{$<k^2_{out}>$}         & \multicolumn{1}{c|}{$<k^2_{in}>$}       & \multicolumn{1}{c|}{$<k^{total}>$}      & \multicolumn{1}{c|}{$<k^{out}>$}       & \multicolumn{1}{c|}{$<k^{in}>$}         & \multicolumn{1}{c|}{$<k^{total}_{std}>$} & \multicolumn{1}{c|}{$Cl.Coef.^{std}_{global}$} & \multicolumn{1}{c|}{$<Cl.Coef.^{undirected}_{local}>$} & \multicolumn{1}{c|}{$<Cl.Coef.^{directed}_{local}>$} \\ \hline
			\multicolumn{1}{|c|}{\textbf{fGn(H=0.3)}}   & \multicolumn{1}{l|}{1539.4886+/-78.7488} & \multicolumn{1}{l|}{385.9506+/-19.7517}    & \multicolumn{1}{l|}{386.1561+/-19.7614} & \multicolumn{1}{l|}{30.5858+/-1.525}    & \multicolumn{1}{l|}{15.2929+/-0.7625}  & \multicolumn{1}{l|}{15.2929+/-0.7625}   & \multicolumn{1}{l|}{3.0424+/-0.15}       & \multicolumn{1}{l|}{0.0194+/-0.001}            & \multicolumn{1}{l|}{0.8481+/-0.0422}                   & \multicolumn{1}{l|}{0.5165+/-0.0258}                 \\ \hline
			\multicolumn{1}{|c|}{\textbf{fGn(H=0.5)}}   & \multicolumn{1}{l|}{1595.5203+/-80.9523} & \multicolumn{1}{l|}{400.1648+/-20.3541}    & \multicolumn{1}{l|}{400.1184+/-20.2721} & \multicolumn{1}{l|}{31.125+/-1.5474}    & \multicolumn{1}{l|}{15.5625+/-0.7737}  & \multicolumn{1}{l|}{15.5625+/-0.7737}   & \multicolumn{1}{l|}{3.1006+/-0.1527}     & \multicolumn{1}{l|}{0.0221+/-0.0011}           & \multicolumn{1}{l|}{0.9208+/-0.0454}                   & \multicolumn{1}{l|}{0.5478+/-0.0272}                 \\ \hline
			\multicolumn{1}{|c|}{\textbf{fGn(H=0.7)}}   & \multicolumn{1}{l|}{1507.9394+/-77.9357} & \multicolumn{1}{l|}{378.3426+/-19.6152}    & \multicolumn{1}{l|}{378.1347+/-19.5056} & \multicolumn{1}{l|}{30.4147+/-1.5221}   & \multicolumn{1}{l|}{15.2073+/-0.761}   & \multicolumn{1}{l|}{15.2073+/-0.761}    & \multicolumn{1}{l|}{2.9856+/-0.147}      & \multicolumn{1}{l|}{0.0173+/-0.0009}           & \multicolumn{1}{l|}{0.8205+/-0.0408}                   & \multicolumn{1}{l|}{0.508+/-0.0251}                  \\ \hline
			\multicolumn{1}{|c|}{\textbf{DJIA-SSEC}}    & \multicolumn{1}{l|}{794.5777}    & \multicolumn{1}{l|}{212.0666}      & \multicolumn{1}{l|}{223.2222}   & \multicolumn{1}{l|}{22.7111}   & \multicolumn{1}{l|}{11.3555}  & \multicolumn{1}{l|}{11.3555}   & \multicolumn{1}{l|}{2.4890}    & \multicolumn{1}{l|}{0.0351}         & \multicolumn{1}{l|}{1.0437}                & \multicolumn{1}{l|}{0.4873}             \\ \hline
			\multicolumn{1}{|c|}{\textbf{DJIA-S\&P500}} & \multicolumn{1}{l|}{140.5777}    & \multicolumn{1}{l|}{37.1777}      & \multicolumn{1}{l|}{34.8666}   & \multicolumn{1}{l|}{10.4444}   & \multicolumn{1}{l|}{5.2222}  & \multicolumn{1}{l|}{5.2222}   & \multicolumn{1}{l|}{0.8365}   & \multicolumn{1}{l|}{0.0140}         & \multicolumn{1}{l|}{0.4092}               & \multicolumn{1}{l|}{0.2770}               \\ \hline
			& \multicolumn{1}{c}{}                     & \multicolumn{1}{c}{}                       & \multicolumn{1}{c}{}                    & \multicolumn{1}{c}{}                    & \multicolumn{1}{c}{}                   & \multicolumn{1}{c}{}                    & \multicolumn{1}{c}{}                     & \multicolumn{1}{c}{}                           & \multicolumn{1}{c}{}                                   & \multicolumn{1}{c}{}                                 \\ \hline
			\multicolumn{1}{|c|}{\textbf{Couplings}}    & \multicolumn{1}{c|}{$Cl.Coef._{global}$} & \multicolumn{1}{c|}{$Sc.Ass.Coef._{var.}$} & \multicolumn{1}{c|}{$<L^{directed}>$}   & \multicolumn{1}{c|}{$<L^{undirected}>$} & \multicolumn{1}{c|}{$R$ from Eq.~\ref{eq1}}                 & \multicolumn{1}{c|}{$Ass.Coef._{var.}$} & \multicolumn{1}{c|}{$Ass.Coef.$}           & \multicolumn{1}{c|}{$Sc.Ass.Coef.$}              & \multicolumn{1}{c|}{$modularity^{total-degree}$}       & \multicolumn{1}{c|}{$modularity^{out-degree}$}       \\ \hline
			\multicolumn{1}{|c|}{\textbf{fGn(H=0.3)}}   & \multicolumn{1}{l|}{0.6983+/-0.0343}     & \multicolumn{1}{l|}{0.0316+/-0.0016}       & \multicolumn{1}{l|}{1.7528+/-0.0862}    & \multicolumn{1}{l|}{1.565+/-0.077}      & \multicolumn{1}{l|}{-0.0258+/-0.0234}  & \multicolumn{1}{l|}{0.0066+/-0.0003}    & \multicolumn{1}{l|}{-0.0049+/-0.0008}    & \multicolumn{1}{l|}{-0.0799+/-0.0079}          & \multicolumn{1}{l|}{-0.0025+/-0.0016}                  & \multicolumn{1}{l|}{-0.0024+/-0.002}                 \\ \hline
			\multicolumn{1}{|c|}{\textbf{fGn(H=0.5)}}   & \multicolumn{1}{l|}{0.7011+/-0.0345}     & \multicolumn{1}{l|}{0.0292+/-0.0014}       & \multicolumn{1}{l|}{1.7269+/-0.0848}    & \multicolumn{1}{l|}{1.5484+/-0.076}     & \multicolumn{1}{l|}{0.0006+/-0.02}     & \multicolumn{1}{l|}{0.0066+/-0.0004}    & \multicolumn{1}{l|}{-0.0035+/-0.0007}    & \multicolumn{1}{l|}{-0.1358+/-0.0117}          & \multicolumn{1}{l|}{-0.0024+/-0.0014}                  & \multicolumn{1}{l|}{-0.0016+/-0.0018}                \\ \hline
			\multicolumn{1}{|c|}{\textbf{fGn(H=0.7)}}   & \multicolumn{1}{l|}{0.688+/-0.0339}      & \multicolumn{1}{l|}{0.0331+/-0.0016}       & \multicolumn{1}{l|}{1.7813+/-0.0874}    & \multicolumn{1}{l|}{1.5886+/-0.0779}    & \multicolumn{1}{l|}{0.0468+/-0.0163}   & \multicolumn{1}{l|}{0.0072+/-0.0004}    & \multicolumn{1}{l|}{0.0008+/-0.0009}     & \multicolumn{1}{l|}{-0.0338+/-0.0087}          & \multicolumn{1}{l|}{0.0072+/-0.0015}                   & \multicolumn{1}{l|}{0.0095+/-0.0019}                 \\ \hline
			\multicolumn{1}{|c|}{\textbf{DJIA-SSEC}}    & \multicolumn{1}{l|}{0.6523}   & \multicolumn{1}{l|}{0.0419}     & \multicolumn{1}{l|}{2.0138}   & \multicolumn{1}{l|}{1.6237}    & \multicolumn{1}{l|}{0.0690} & \multicolumn{1}{l|}{0.0102}   & \multicolumn{1}{l|}{0.0006}   & \multicolumn{1}{l|}{-0.1917}        & \multicolumn{1}{l|}{-0.0052}                & \multicolumn{1}{l|}{-0.0197}              \\ \hline
			\multicolumn{1}{|c|}{\textbf{DJIA-S\&P500}} & \multicolumn{1}{l|}{0.5221}   & \multicolumn{1}{l|}{0.0411}     & \multicolumn{1}{l|}{3.3819}   & \multicolumn{1}{l|}{3.4666}   & \multicolumn{1}{l|}{0.3414} & \multicolumn{1}{l|}{0.0297}  & \multicolumn{1}{l|}{0.1491}   & \multicolumn{1}{l|}{0.6630}         & \multicolumn{1}{l|}{0.0118}                 & \multicolumn{1}{l|}{-0.0304}              \\ \hline
	\end{tabular}	}  
	\caption{To test the reliability and robustness of the results, we estimated
		the mean interval of each measurement for the network obtained from fGns.
		The estimated interval for each measurement is calculated with the
		confidence level of $90\%$ by$\bar{x}-Z_{(1-\frac{\protect\alpha}{2}%
			=0.95)}\times\frac{S}{\protect\sqrt{n}} <\protect\mu < \bar{x} +Z_{(1-\frac{%
				\protect\alpha}{2}=0.95)}\times\frac{S}{\protect\sqrt{n}}$; where $\protect%
		\mu$ is the measurement value. $\bar{x}$, $S$ and $n$ stand for the sample
		mean, the sample standard deviation and the number of generated samples,
		respectively. As shown, the results in the table conform with Fig~\protect
		\ref{fig_result}. Some measurements in the DJIA-SSEC mapped network and the
		DJIA-S\&P500 mapped network have similarity with the measurements generated
		by fGns and some have larger deviations.}
	\label{tab:1}
\end{table*}
where, $\sigma $ denotes the standard deviations along the diameters of the
joint probability matrix. The relationship between the parameter $R$
relative to the corresponding Hurst exponents is shown in Fig.~\ref%
{network_topology}.
\section{Mapping Coupled Time-series Onto a Network}
We map the coupling of two market time-series onto a network. The algorithm
which is applied here, is the same as the previous one. However, two
time-series with the simultaneous chronological time-stamp (no lag) are
considered. The outcome will be compared with the fGns which are already
mapped onto the network (Fig.~\ref{fig_result}). In Fig.~\ref{fig_algo}, in
addition to considering the positive and negative amplitudes, we account for
the differences between the amplitudes. The placement of amplitudes in the
same amplitude-bin, leads to a self-loop. On the other hand, the placement
of amplitudes in different amplitude-bins leads to an edge. The direction of
edges stands for emphasizing the difference between whether the first signal
is in bin A and the other one in B, as opposed to whether the first signal
is in bin B and the second one in A.

\section{Results and Discussion}

In Fig.~\ref{fig_result}, radar plots a) and b), 20 topological and
statistical measurements of the obtained networks from the cross-markets are
compared with those from fGns and those from surrogate time-series. It is
notable that a fGn with Hurst=0.5 is the indication of no coupling. The
convergence of any cross-market measurement to the measurement related to
Hurst=0.5, illustrates insignificant information embedded in the coupling.
Despite the segregation among the measurements of different joint systems,
there exist some similarities. As shown in Fig.~\ref{fig_result}, the
DJIA-SSEC's coupling is closer to an uncoupled situation rather than
DJIA-S\&P500's coupling. In radar plot b), the same measurements are
reported for the joints of surrogate DJIA vs. surrogate SSEC, and also,
for the joints of surrogate DJIA vs. surrogate S\&P500. A non-Gaussian
time-series gives up its non-Gaussianity by surrogate method. Hence, it does not
have any effect on the fGns. The sources of different coupling between
time-series stem from two phenomena: correlation, and fat-tailed
distribution. After surrogate method, the correlation remains, but the
probability distribution converts to a Gaussian distribution. To consider
more extensive test systems, in radar plot b), the market series are
surrogated (see Appendix A) and the coupling of surrogate series were
mapped to a network. The outcome is interesting where the patterns have
higher conformity rather than the original time-series in radar plot a).
Radar plot b) shows that, after surrogate method, the deviations which their
sources relate to the fat-tailed PDFs, come closer to each other. It means
that the deviations with the source of non-Gaussianity are eliminated, and
the deviations with merely the source of correlation remain.\newline
\textit{- Deformation Ratio (R):} In Fig.~\ref{fig_result} a comparison
between joint probability matrices of DJIA-SSEC and DJIA-S\&P500 is shown.
The strength of couplings are visually shown. The coupling of DJIA-S\&P500
is stronger than that of DJIA-SSEC. This feature is quantified based on Eq.~%
\ref{eq1} with the $R$ parameter which is considered in the radar plots in
Fig.~\ref{fig_result}.\newline
\textit{- Degree Measurements:} The measurements corresponding to degrees,
such as mean squared out-degrees $<k_{out}^{2}>$, mean squared in-degrees~$%
<k_{in}^{2}>$, mean squared total-degrees~$<k_{total}^{2}>$, mean
out-degrees~$<k_{out}>$, mean in-degrees $<k_{in}>$ and mean total-degrees~$%
<k_{total}>$, contain significant power for proving the segregation among
cross-markets and the fGn with Hurst=0.5. The mentioned features in the
radar plots in Fig~\ref{fig_result} contain significant coupling information
among fGns and also the cross-market coupling mapped onto a network. Also,
the standard deviation of total-degree~$<k_{std}^{total}>$ turns up to
identify cross-market couplings.\newline
\textit{- Clustering Measurements~\cite{Watts1998}:} The standard deviation
of the global clustering coefficient~$Cl.Coef._{global}^{std}$ is capable of
exploring the difference between cross-market couplings. The undirected
local clustering coefficient~$Cl.Coef._{local}^{undirected}$ can distinguish
among the networks of coupled and uncoupled cross-markets. This feature
converges to a fGn with Hurst=0.5 for the network extracted from an
uncoupled cross-market. Also, the directed local clustering coefficient ~$%
Cl.Coef._{local}^{directed}$ is different for uncoupled and coupled
outcomes. The global clustering coefficient~$Cl.Coef._{global}$ for fGns,
uncoupled and coupled cross-markets are approximately similar.\newline
\textit{- Length (Shortest Path Between Pair-wise Vertices) Measurements:}
It is striking that the directed mean length~$<L^{directed}>$ and the
undirected mean length~$<L^{undirected}>$ significantly explore the
differences between coupled cross-markets from uncoupled cross-markets and
the fGns.\newline
\textit{- Assortativity Measurements~\cite{Newman2003}:} The variance of
scalar assortativity coefficient~$Sc.Ass.Coef._{var.}$ for fGns, uncoupled
and coupled cross-markets are approximately similar. Conversely, the
assortativity coefficient variance~$Ass.Coef._{var.}$, assortativity
coefficient~$Ass.Coef.$, scalar assortativity coefficient~$Sc.Ass.Coef.$
markedly distinguish among coupled cross-markets from fGns and uncoupled
cross-markets.\newline
\textit{- Modularity Measurements~\cite{Newman2006,NewmanGirvan2004}:} As
shown in Fig.~\ref{fig_result} notwithstanding that out-degree modularity
enables one to identify the cross-markets from fGn, the total-degree
modularity~$Modularity^{total-degree}$ is highly capable of showing the
divergence between uncoupled and coupled cross-markets. Based on the
out-degree modularity measurement, there exists mutual information among
markets and it is not uncoupled in this manner. In this regard, markets are
coupled or weakly coupled (not necessarily uncoupled).\newline
To further assess the patterns, DJIA-S\&P500's coupling is adequately far
from white noise (a fGn with Hurst = 0.5) and DJIA-SSEC's coupling. The
flipside of the coin is that DJIA-SSEC's coupling is closer to white noise
(an fGn with Hurst = 0.5) rather than DJIA-S\&P500, but it is still totally
different. Although the joint probabilities of DJIA and SSEC's time-series
show that they are uncoupled (Eq.~\ref{eq1} and Fig.~\ref{fig_result}), by
mapping the coupling of two time-series onto a network, more hidden
properties are revealed. According to some other network measurements, those
markets still possess coupling information. Hence, being a market
contributes to being coupled with others. Thus, it is better to use the
term, \textit{weakly coupled markets} rather than the term, \textit{%
	uncoupled markets}. Along with giving us the ability to measure the coupling
constituents between two time-series, Fig.~\ref{fig_result} will extend our
knowledge toward realistic simulations in joint structures within a network
perspective.\\

	Financial shocks contribute to contagion through alteration in couplings and dependencies~\cite{Medovikov2017}.
	Our results by mapping the couplings~\cite{Ardalankia2020,Caraiani2015} onto a network have applications 
	in diverse risk measurement approaches with internal and multilateral interactions in economic and financial networks~\cite{Battiston2016,Perillo2018},
	and crisis analysis
	~\cite{Amini2010,Habibnia2017,Kaushik2013,Shirazi2017}.
	If an economist can distinguish where the sources of couplings stem from, 
	and under what circumstances the prices change simultaneously, 
	they can find out how to conform an investment portfolio so to lower the risk. 
	It is of great importance to know how couplings are formed from the PDFs and their tails. 
	To explain it more in details we have:\\
	- Degree Measurements: The tails of the PDF possess lower
	degrees, and, the nodes near the mean value possess higher degrees.
	$<k>^2$ contains more information about the tails
	as opposed to $<k^2>$.
	Conversely, $<k^2>$ contains more information about the mean value as opposed to $<k>^2$.
	Hence, the relation $\frac{<k>^2}{<k^2>}$ contains vital information about the source of coupling, i.e. whether the coupling stems from a fat tail distribution or the Gaussian distribution.
	High $\frac{<k>^2}{<k^2>}$ means that tails contain a higher share in the couplings.
	Low $\frac{<k>^2}{<k^2>}$ implies that the tails contain a smaller share in the couplings.\\
	- Assortativity: A high assortativity means that a high-degree component 
	usually create links with the high-degree ones, and vice versa. It implies the system tends to preserve its trend. 
	The disassortativity quantity implies that a big jump tends to a small movement.\\
	- Modularity: Each node is a price return. Modularity shows that the changes within a community are more probable than between two communities. It can be assessed whether the modularity is a mathematical language for the terms ``support'' and ``resistence'' in technical market analysis. High modularity implies that the corresponding price returns of the two markets follow simultaneous similar changes.
	\begin{figure}[h]
		\centering
		\includegraphics[width=0.3\textwidth]{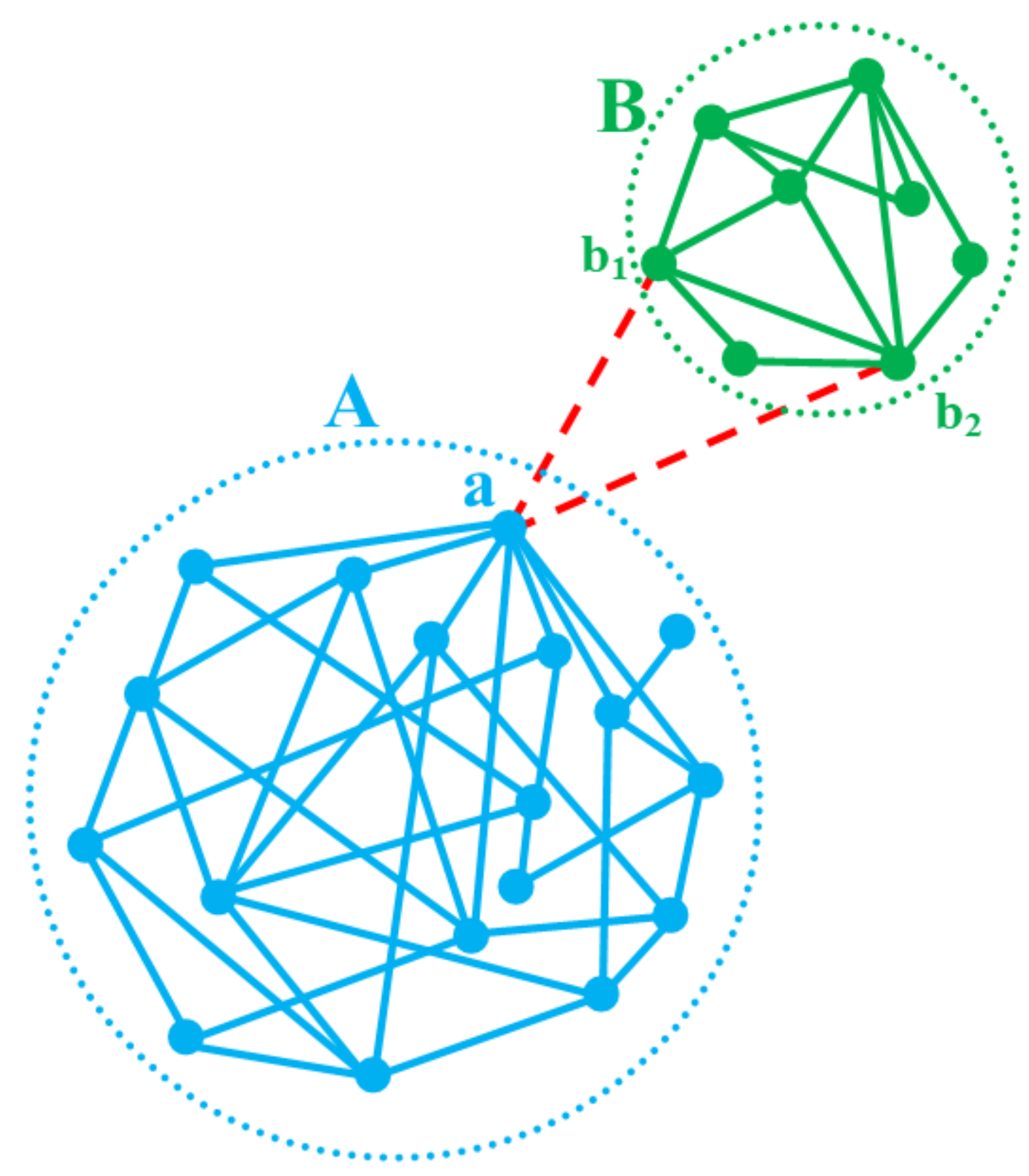}
		\caption{This figure schematically shows two communities (A and B). There are certain nodes (price returns) that can connect both communities via those nodes (price returns) --$a$ to $b_1$ and $b_2$ (via red dashed lines). Inter-community movements from one node (price return) to another node (another price return) is more probable, and intra-community movements are less probable. Thereby, higher modularity leads to lower homogeneouty in the likelihood of movements throughout the whole network. Indeed, modulaity quantifies the heterogeneouty of the possibility of the inter-community and intra-community of movements.}
		\label{figCommunity}
	\end{figure}\\
	- Mean Length: It declares that on average how agents (say amplitudes in our case) 
	can create a relation with each other. In other words, 
	it shows to what extent the system can translate its dynamic.\\
	- Deformation Ratio: It is applied to quantify the joint probability shape. 
	Based on the deformation ratio, one can find out how two joint-markets are extended 
	in comparison with an uncoupled market.\\
	- Clustering Coefficient: A high clustering coefficient states the extent to which the agents in the system 
	tend to remain in their clusters. In our case, the amplitudes resist to change their clusters.\newline
Even two previously known uncoupled and uncorrelated markets may possess
coupled characteristics. Hence, those markets should be examined as coupled
and \textit{weakly} coupled markets.
\section{Conclusion}
Mapping the cross-correlation of two coupled time-series onto a network
helps scholars to gain more insight into the important constituents of joint
structures between two time-series. Topological and statistical parameters
along with the \textit{deformation ratio of joint probability between two
	time-series} (which is extracted from the standard deviations along both
diameters of the directed weighted adjacency matrix) are able to reveal the
coupling information which has previously been beyond the reach of
researchers. Comparing the network mapped from fGns and from joint markets
structures, not only proves pair-wise inter-connectedness, it also clarifies
the diverse structure of the coupling and cross-correlation therein. The
reasoning behind this claim is that couplings with different Hurst exponents
show a diverse range of behaviours (anti-persistent, white noise,
persistent). Those behaviours can be reflected in a network from mapping the
joint structures to that network. Also, the network mapped from joint
structures of surrogate time-series in Fig.~\ref{fig_result} proves that
the coupling is derived from two criteria: cross-correlation, and a
fat-tailed PDF. 
\section{Appendix A}
The surrogate method converts the non-Gaussian PDF to the Gaussian PDF. Although surrogate method eliminates the nonlinear structure in the time-series, it maintains the linear structure. The outcome gets closer to a Gaussian process.
	Through a Fourier surrogate, after the phase randomization process, the central limit theorem is satisfied.
	Given a time-series named $X(t)$, the discrete Fourier transform of $X(t)$ is given by:\\
	\begin{equation}
	X(\omega)=\frac{1}{N}\sum_{t} X(t) \exp(i \omega t);\\
	\label{eq2}
	\end{equation}
	Then, the phase of the time-series is randomized by a pseudo-independent uniform distribution set, $\eta$. 
	Thereby, we have:
	\begin{equation}
	X^{*}(\omega)=\sum_{\omega} |X(\omega)| \exp(-i \eta \omega);\\
	\label{eq3}
	\end{equation}
	Since sine and cosine values in Eq.~\ref{eq2} occur within [-1,1], $X_{max}(\omega)\leq X_{max}(t)$, 
	and $x(t)\neq \infty$, then $X(\omega)$ has finite mean and variance. 
	Hence, based on the central limit theorem in a randomization procedure, the PDF translates to a Gaussian PDF.
	Accordingly, by applying a reverse discrete Fourier transform on $X^{*}$, 
	the resulting phase randomized time-series is Gaussian~\cite{Hedayatifar2011}.

\bibliographystyle{elsarticle-num}
\bibliography{arXiv.bib}

\end{document}